\documentclass[runningheads]{llncs}
\usepackage[T1]{fontenc}
\usepackage[a4paper, margin=1in]{geometry}
\usepackage{graphicx}
\usepackage{hyperref}
\usepackage{booktabs}
\usepackage{caption}
\usepackage{multirow}
\usepackage{rotating}
\usepackage{forest}
\usepackage{tabularx}
\usepackage{array}
\newcolumntype{Y}{>{\raggedright\arraybackslash}X}

\newcommand\blfootnote[1]{%
  \begingroup
  \renewcommand\thefootnote{}%
  \footnotetext{#1}%
  \endgroup
}

\begin{document}

\title{Signal-based Model Access Risk Analysis for AI System Operations Security}
\author{Maria Mahbub\inst{1,*} \and
Steven Young\inst{1} \and
Amir Sadovnik\inst{1} \and
Edmon Begoli\inst{1} \and
Chris Rugenstein\inst{2} \and
Donald Coulter\inst{3} \and
Anthony Ayodele\inst{3}} 

\authorrunning{M. Mahbub et al.}

\institute{Oak Ridge National Laboratory, Oak Ridge, TN 37830, USA\\
\email{\{mahbubm,youngsr,sadovnika,begolie\}@ornl.gov}\\
\and
Oak Ridge Institute for Science and Education, Oak Ridge, TN 37830, USA
\email{christopher.rugenstein@associates.hq.dhs.gov}
\and
DHS Science and Technology Directorate, Washington, DC 20528, USA\\
\email{\{donald.coulter,anthony.ayodele\}@hq.dhs.gov}}


%
\maketitle              

\begin{abstract}

Artificial intelligence (AI) systems are now ubiquitous across domains such as security, finance, healthcare, consumer technology, and large-scale cloud services, where they process massive volumes of data and make consequential decisions daily. This widespread adoption has created a broad attack surface through which adversaries can manipulate, evade, extract information from, or otherwise subvert deployed models. Depending on system design and exposure, attackers may have very different forms of access: some observe only final decisions, while others receive confidence scores, intermediate representations, or even full model parameters. While previous surveys typically organize evasion attacks into white-box, gray-box, and black-box categories based on the attacker's knowledge of model internals (architecture, parameters, gradients), this taxonomy often conflates different deployment scenarios that provide vastly different output signals, all labeled as ``black-box'' despite enabling fundamentally different attack strategies. Understanding how evasion attack strategies adapt to the specific information signals returned by deployed systems is critical for organizations making procurement and deployment decisions.
To address this gap, we introduce the Signal-based Model Access Risk Taxonomy (SMART), a deployment-oriented framework that classifies attacker access according to the nature and richness of the information signals available from deployed AI systems. Using this taxonomy, we provide a structured overview of evasion attacks across progressively richer levels of information exposure, highlighting how deployment interfaces influence attack capabilities and informing more secure AI deployment and procurement decisions.


\end{abstract}

\section{Introduction}
\blfootnote{Notice: This manuscript has been authored by UT-Battelle, LLC, under contract DE-AC05-00OR22725 with the US Department of Energy (DOE). The US government retains and the publisher, by accepting the article for publication, acknowledges that the US government retains a nonexclusive, paid-up, irrevocable, worldwide license to publish or reproduce the published form of this manuscript, or allow others to do so, for US government purposes. DOE will provide public access to these results of federally sponsored research in accordance with the DOE Public Access Plan (\url{https://www.energy.gov/doe-public-access-plan}).}

Artificial intelligence (AI) systems have become integral to decision-making across domains including healthcare, finance, cybersecurity, autonomous systems, and cloud-based services. As these models are increasingly exposed through public APIs, enterprise platforms, and consumer applications, they face growing risks from adversarial evasion attacks that manipulate model inputs to induce incorrect predictions without altering the underlying model. The security of these deployed systems depends not only on the robustness of their learning algorithms but also on the information made available through their deployment interfaces.

In practice, deployed AI systems expose markedly different levels of information to potential adversaries. Some systems reveal only a final decision, whereas others return confidence scores, ranked predictions, feature embeddings, or even provide direct access to model parameters. These differences fundamentally shape the attack strategies available to an adversary, influencing both the feasibility and effectiveness of evasion attacks. However, existing surveys predominantly classify evasion attacks using knowledge-based categories such as white-box, gray-box, and black-box, which often group together deployment scenarios that expose substantially different information signals. As a result, the relationship between deployment interface design and evasion risk remains insufficiently characterized.

We introduce a Signal-based Model Access Risk Taxonomy (SMART), which organizes evasion attack scenarios by the nature and richness of information signals accessible to an adversary across six access categories: None, Metadata, Decision-Only, Score/Rank, Embedding, and White-Box.
By analyzing how progressively richer access levels transform the evasion vulnerability landscape across AI system pipelines, this taxonomy provides organizations, evaluating AI model procurement options from building in-house solutions to contracting commercial vendors, with a structured framework for understanding how evasion risk scales with exposure.
The analysis demonstrates that evasion threats exist across all access levels and shows how architectural choices, API-based versus on-premise, open-source versus proprietary, cloud versus edge deployment, map to specific vulnerabilities, supported by procurement decision frameworks, research references, and security evaluation criteria for informed risk assessment during AI system selection and deployment.
To ground the taxonomy in a concrete setting, the analysis is demonstrated using face recognition systems, which are widely deployed in real-world applications including airport security, mobile devices, access control, surveillance infrastructure, and cloud-based vision services, creating numerous attack surfaces.

Throughout this paper, we will utilize face recognition systems as a motivating example, but the taxonomy can be utilized for any AI system. Face recognition systems are widely used in governmental, commercial, and consumer applications, ranging from border control to smartphone unlock mechanisms and financial authentication. As organizations increasingly deploy these systems for security-critical functions, understanding their vulnerability to evasion attacks, adversarial inputs designed to cause misclassification or impersonation, becomes essential for risk management and procurement decision-making.
The security posture of a face recognition system is not solely determined by the underlying model's accuracy or the vendor's reputation. Rather, it is fundamentally shaped by the level of access an adversary can obtain to the system. As AI systems become ubiquitous, it is important to systematically evaluate the risk emanating from acquisition and deployment choices as a result of the level of access provided to attackers.



We address this gap by presenting SMART, a signal-based access taxonomy, that categorizes adversary access along a spectrum from no knowledge of the AI system to complete system knowledge. For each access category, we analyze examples of attack methodologies documented in peer-reviewed research, if available, discuss attack efficiency in terms of query budgets and success rates, examine the procurement scenarios that lead to each access level, and provide evaluation criteria for organizations selecting among deployment options.

SMART is particularly relevant for organizations facing the following procurement decisions:
\textit{(1) Build versus buy:} Should we develop in-house face recognition capabilities, leverage open-source models, or procure commercial systems?
\textit{(2) Deployment architecture:} Should we deploy on-premise, use cloud-based APIs (Azure Face API, AWS Rekognition, Google Cloud Vision), or implement edge-based solutions?
\textit{(3) Access control design:} Should our system return binary decisions, confidence scores, or expose embeddings for interoperability?
\textit{(4) Vendor selection:} How do we evaluate security claims from commercial vendors, and what contractual protections should we require?
\textit{(5) Custom development:} When contracting developers to build proprietary systems, what security requirements and architectural constraints should we specify?

Organizations increasingly face these choices with limited guidance on how deployment decisions affect adversarial robustness. A common misconception is that restricting external access or using proprietary closed-source systems provides adequate security. However, research demonstrates that sophisticated attacks succeed even against systems with minimal observable signals, while open-source deployments and certain commercial architectures operate under effectively white-box threat models.

The taxonomy presented here synthesizes over a decade of adversarial machine learning research, with particular emphasis on face recognition-specific vulnerabilities. Each access category is supported by citations to seminal research papers that establish attack feasibility, efficiency, and transferability. By mapping procurement options to access categories and access categories to attack surfaces, this framework enables evidence-based security evaluation during system selection.

The remainder of this paper is organized as follows: Section \ref{sec:taxonomy} presents the entire access taxonomy with definitions and example deployment scenarios. It then analyzes each access category in detail, covering attack methodologies, key research papers, and procurement evaluation criteria. Section \ref{sec:decision} provides a comprehensive risk landscape analysis and procurement decision considerations, and then concludes with strategic recommendations for organizations procuring face recognition systems in adversarial environments.

\section{Signal-Based Model Access Risk Taxonomy}
\label{sec:taxonomy}

\begin{figure}[!htbp]
    \centering
    \includegraphics[width=\linewidth]{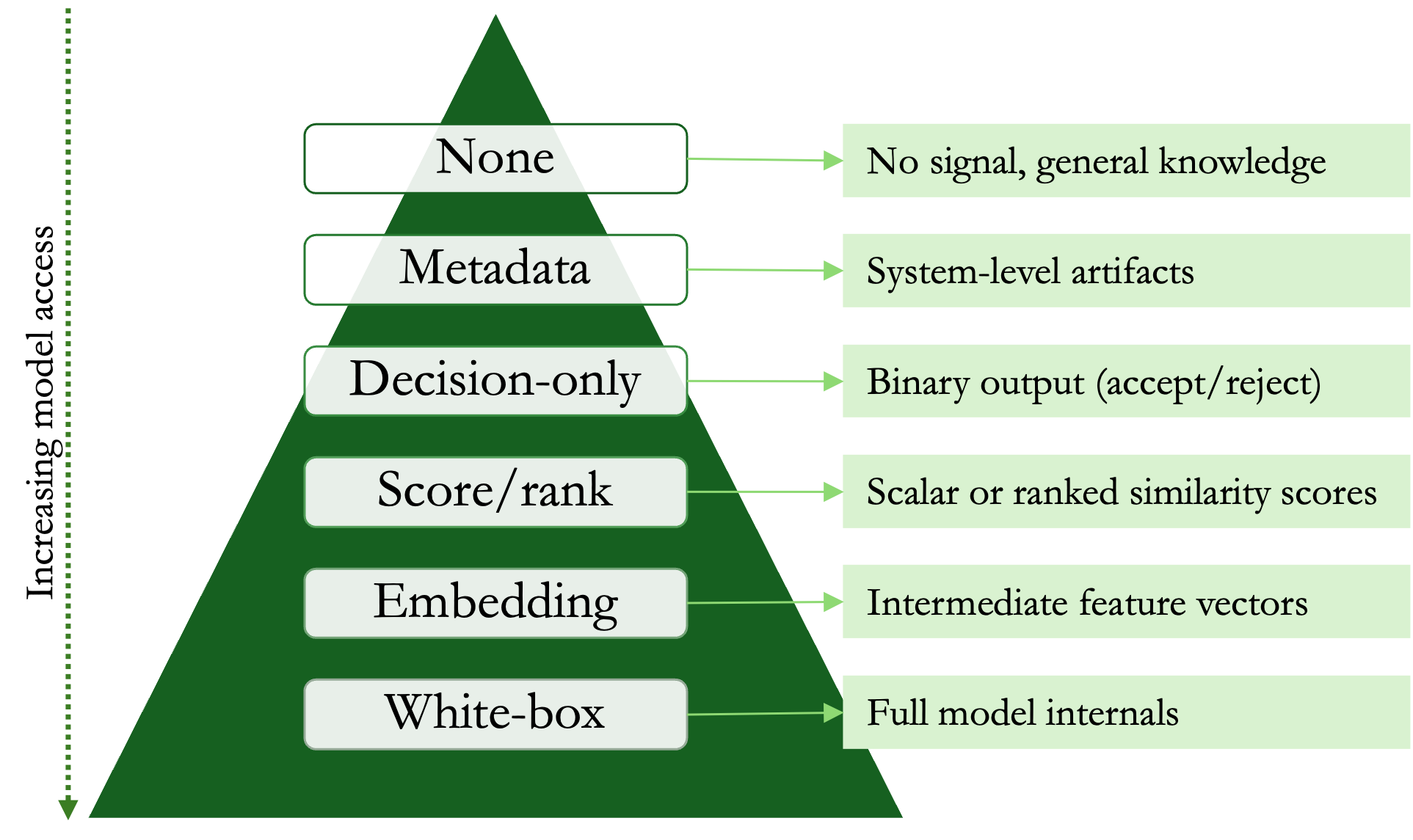}
    \caption{Signal-based Model Access Risk Taxonomy (SMART)}
    \label{fig:access-taxonomy}
\end{figure}

Figure \ref{fig:access-taxonomy} and Table \ref{tab:access-taxonomy} present the complete taxonomy of access levels, organized by the type of signal or information available to the adversary. This classification scheme progresses from minimal knowledge scenarios (None) through increasingly informative feedback mechanisms (Decision-Only, Score/Rank, Embedding) to complete model transparency (White-Box). Each category represents a qualitatively different threat model with distinct attack methodologies and efficiency characteristics.

\begin{table}[!htbp]
\centering
\caption{Signal-based model access risk taxonomy with access descriptions and example scenarios.}
\label{tab:access-taxonomy}
\begin{tabularx}{\textwidth}{p{2cm}p{2cm}YY}
\hline
\textbf{Access Category} & \textbf{Signal Type} & \textbf{Access Description} & \textbf{Example Systems / Scenarios} \\
\hline

\textbf{None} &
No signal exposure &
The adversary has general knowledge of the system but no details on the model being used. &
Attacker knows organization uses face recognition systems at security checkpoints. \\
\hline

\textbf{Metadata} &
System-level artifacts &
Access to metadata from sales brochures, technical data sheets, etc. that give limited information about the model or access limited to observable side channels such as latency, power, or error codes, which can leak model information. &
Technical data sheet specifying RAM, compute power. Company owner's dissertation on face recognition models. \\
\hline

\textbf{Decision-Only} &
Binary output (accept/reject) &
Only verification outcomes are visible; attacker can query inputs and observe yes/no decisions. &
Device unlock, access control, gatekeeping systems, some API systems. \\
\hline

\textbf{Score/Rank} &
Scalar or ranked similarity scores &
Outputs include confidence levels or similarity values, enabling approximation of model boundaries or templates. &
Most commercial APIs (e.g., Face++, AWS Rekognition) providing score-based matching. \\
\hline

\textbf{Embedding} &
Intermediate feature vectors &
Access to internal representations used for downstream comparison or clustering; reveals high-dimensional feature space. &
Interoperable face/vision SDKs sharing embeddings across systems. Some commercial APIs. \\
\hline

\textbf{White-Box} &
Full model internals &
Complete access to parameters, architecture, and gradients; enables model replication and inversion. &
Open-source model leaks, insider access, reverse engineering or collaborative training environments. \\
\hline
\end{tabularx}
\end{table}

The taxonomy deliberately focuses on the information available to the adversary rather than the specific deployment architecture, because the same deployment choice may result in different access levels depending on operational context and adversary capabilities. For example, an on-premise face recognition system might provide Decision-Only access to external attackers, Metadata access to supply chain adversaries who obtain technical documentation, and White-Box access to insider threats with physical device access.

Understanding this taxonomy is critical for acquisition, as traditional cybersecurity and performance metrics don't capture adversarial risk. Selecting a commercial API for convenience may unintentionally provide Score/Rank access to any internet-connected adversary. Deploying open-source models for cost savings guarantees White-Box access from day one. Choosing SDK-based architectures for interoperability may expose Embeddings that enable both evasion and privacy attacks. Any of these may be suitable choices, but their impact on the specific system being designed must be understood.

\subsection{None: No Knowledge, No Interaction, No Signal}
In the None access category, the adversary has no actionable knowledge of the deployed system beyond the fact that the organization uses an AI-based face recognition (FR) capability. The attacker does not know the vendor, architecture, training data, deployment modality, or operational thresholds, and cannot interact with or query the system in any way. This represents the weakest possible attacker in terms of access and is often implicitly assumed to pose negligible risk.

Despite this, evasion risk is not necessarily eliminated. An example is modern face recognition where systems share structural and statistical properties that can be exploited without targeting a specific model. Research on adversarial examples has repeatedly shown that perturbations crafted for one model frequently transfer to other models trained for the same task, even when architectures and training datasets differ \cite{goodfellow2014explaining,zhong2020towards,papernot2017practical}. This phenomenon of transferability allows attackers to operate without any knowledge of the victim system.

The transferability phenomenon underlying no-access attacks can be partially explained by the shared learning objectives of face recognition models. Most modern systems are trained using metric learning approaches (triplet loss, ArcFace, CosFace) that optimize for similar embedding space geometries. When adversaries craft perturbations that exploit these common geometric properties, for example by pushing embeddings across decision boundaries in high-dimensional space, the perturbations often transfer to other models optimized for similar objectives, even when implemented with different architectures. This structural similarity across face recognition systems creates a baseline vulnerability that exists independent of deployment-specific defenses.
Research demonstrates that adversarial perturbations generated on surrogate models can successfully degrade performance on unseen target systems without requiring queries, observations of outputs, or knowledge of model internals. Multiple studies have shown that attackers can rely solely on public datasets and commonly used architectures to generate inputs that evade recognition with non-trivial probability, including impersonation and obfuscation attacks against state-of-the-art face recognition systems and commercial APIs \cite{deb2020advfaces,zhong2020towards}.

A second evasion pathway under the None category is physical-world presentation attacks, which target the input acquisition process rather than the classifier itself. These attacks exploit weaknesses in face capture, alignment, and normalization by using printed artifacts, adversarial patches, or wearable accessories. That is, these attacks don't directly attack the model, but the common pre-processing steps. Researchers demonstrate that physically realizable adversarial eyeglass frames can cause targeted misclassification in face recognition systems, with success rates exceeding 80\% in controlled settings \cite{sharif2016accessorize}. Importantly, these attacks generalize across different face recognition systems because they exploit consistent pre-processing assumptions, face detection, alignment, normalization, that are common to most commercial implementations \cite{vakhshiteh2021adversarial,wang2025survey}.

Organizations should consider these risks when evaluating whether to build custom solutions versus deploying well-known architectures. Even attackers with no system knowledge can exploit shared vulnerabilities of visual recognition pipelines. While novel or proprietary architectures might provide some security through obscurity, this should never be the primary defense. Procurement decisions must account for the need to defend against transfer attacks regardless of access controls. Many mitigations at this level are available, relying on input validation, anomaly detection, adversarial robustness training, environmental controls, and operational safeguards such as human oversight. The cost-benefit analysis should recognize that even the most restrictive deployment model (air-gapped, no external access) does not eliminate evasion risk, particularly when the proprietary system is replicating a commonly available, open-source capability, e.g. AI-based face recognition.

\subsection{Metadata: Coarse System Knowledge and Low-Resource Observation}
The Metadata access category assumes an adversary who gains coarse, publicly observable information about the system through marketing materials, compliance documentation, high-level system descriptions, or simple operational observations such as approximate response latency. The attacker does not possess privileged access, does not perform sophisticated side-channel analysis, and does not receive explicit model outputs such as scores or embeddings.

Even this limited information meaningfully reduces attacker uncertainty. Knowledge of whether inference is performed on-device or in the cloud, whether the system supports 1:1 verification or 1:N identification, or whether it incorporates liveness checks allows the attacker to narrow the plausible design space. While the literature on evasion attacks has yet to study the effect of metadata, the model stealing literature establishes that architectural knowledge improves surrogate model effectiveness, as prior knowledge about model architecture family reduces the computational cost of training effective surrogates \cite{tramer2016stealing,oh2019towards}. While attackers in the metadata category lack query access, publicly disclosed architectural information, such as statements that a system uses ``convolutional neural networks for facial recognition'' or ``lightweight models optimized for mobile deployment'', enables more focused surrogate selection. An attacker who knows the target uses mobile-optimized architectures can focus training on models like MobileNet or EfficientNet rather than exploring the full architectural space.

However, the practical impact of metadata alone on evasion attack is highly dependent on the metadata available. Transfer-based attacks without queries or outputs from the target system achieve success rates that vary widely depending on architectural similarity between surrogate and target. Publicly available metadata for commercial models typically provides insufficient detail to guarantee architectural alignment, and architectural mismatches significantly degrade attack effectiveness. The Metadata category represents an elevated threat compared to the None category, but the advantage is primarily in reducing attacker costs rather than substantially improving attack success rates. For operational security, organizations should recognize that technical documentation, performance claims, and architectural descriptions, while individually benign, cumulatively enable more efficient attacker strategies, even absent direct system access.

\subsection{Decision-Only: Binary Accept/Reject Feedback}
In the Decision-Only access category, the adversary submits inputs to the system and observes only binary verification outcomes, match/no-match or accept/reject. No additional information is provided: no confidence scores, no similarity values, and no internal representations. This access level characterizes deployment scenarios prioritizing privacy and operational security, including smartphone unlock mechanisms, physical access control gates, and certain API implementations designed to minimize information leakage. Decision-only interfaces are frequently assumed to offer strong protection by withholding detailed system outputs. In practice, however, this protection is incomplete \cite{brendel2017decision,chen2020hopskipjumpattack,dong2019efficient}. Repeated binary queries allow attackers to systematically probe where the system draws the line between ``accept'' and ``reject'', gradually mapping out this acceptance boundary. In face verification specifically, each binary outcome reveals whether two faces are similar enough to exceed a hidden internal threshold. Attackers exploit this signal by repeatedly testing variations of faces until they understand what causes the system to flip its decision.

Empirical studies demonstrate that commercial face recognition APIs can be compromised through adversarial attacks that enable both targeted and untargeted impersonation using only match/no-match outputs \cite{dong2019efficient,chen2020rays,li2020qeba}. Prior work shows that iterative query refinement can approximate a model's decision boundary sufficiently to synthesize effective adversarial inputs, even under decision-only access. Attack effectiveness increases when adversaries obtain or construct surrogate models of the target system; however, these surrogates need not replicate internal model details to be useful, as approximating the decision boundary alone suffices to generate transferable perturbations \cite{orekondy2019knockoff,correia2018copycat,yu2020cloudleak}. With such surrogates, adversaries can perform extensive offline optimization and robustness evaluation, then deploy attacks against live systems with minimal queries, effectively mitigating the constraints imposed by limited-access threat models.
Digital evasion methods leverage this offline-online strategy. Decision-based attacks iteratively perturb pixels until labels flip; GAN-driven methods synthesize realistic faces inducing misclassification; localized patch attacks insert small, high-impact patterns robust to image transformations \cite{goodfellow2014explaining,brown2017adversarial,truong2021data}. These techniques, optimized for transferability in surrogate models, retain effectiveness when deployed against decision-only targets with constrained query budgets.


From a procurement and governance perspective, the actual risk is twofold: (i) decision-only interfaces reduce information leakage but still admit boundary inference with sufficient queries, and (ii) transferable attacks, digital or physical, let adversaries do most work offline using few decision-only trials to succeed. Mitigations to this risk include those from the "None" category along with strict telemetry/rate-limits on verification endpoints and agreements with API providers to limit access to trusted entities or monitor for system probing.

\subsection{Score/Rank: Continuous Confidence and Similarity Feedback}
In the Score/Rank access category, the adversary receives continuous scalar outputs representing confidence levels, similarity scores, or ranking information in response to queries. This access level is the prevailing paradigm for commercial face recognition APIs, including AWS Rekognition, Microsoft Azure Face API, Google Cloud Vision API, and Face++, which return calibrated similarity scores to enable application-level decision thresholds. The transition from binary decisions to continuous scores fundamentally alters the attack surface by enabling gradient estimation techniques that dramatically reduce the query complexity of adversarial example generation.

The core advantage conferred by score access is the ability to approximate gradients through finite difference methods. Rather than requiring thousands of queries to probe decision boundaries discretely, an attacker can estimate the local gradient direction by observing how small input perturbations affect the output score. This converts the adversarial optimization problem from a discrete search over the binary decision space to a continuous optimization problem amenable to established numerical methods. The query efficiency improvement is typically two to three orders of magnitude: attacks requiring 20,000+ queries under decision-only access often succeed with 200–500 queries when score feedback is available \cite{chen2017zoo,tu2019autozoom}.

Researchers have demonstrated practical attacks exploiting score-based feedback \cite{papernot2017practical,dong2019efficient,sharif2016accessorize}. Studies show that by systematically querying systems and observing score changes, attackers can craft adversarial faces that fool commercial APIs with surprisingly few queries. The general approach involves making small modifications to face images, observing how similarity scores respond, and using this information to guide further modifications toward the attack goal. Modern techniques combine this direct querying with offline preparation: attackers train their own models using publicly available face recognition systems, develop candidate attacks against these models, then use a limited number of queries to the target commercial API to fine-tune their attack \cite{ilyas2018black,ilyas2018prior}. This hybrid strategy achieves success rates exceeding 90\% against commercial services while staying within modest query budgets of a few hundred requests, small enough to evade simple rate limiting but sufficient to reliably compromise the system.

Beyond direct attacks, score-based access creates vulnerability to model extraction, where adversaries systematically query the system to build their own copy of it \cite{tramer2016stealing}. The query requirements vary significantly by model complexity: simpler models like logistic regression can be extracted with just dozens to hundreds of queries, while neural networks may require thousands of queries. Once extracted, the copied model becomes a perfect testing ground for developing adversarial attacks, enables privacy violations by reconstructing training data, and may constitute intellectual property theft. For organizations purchasing commercial face recognition services, this represents compounding risk: the same score-based interface that enables direct attacks also facilitates model theft, which in turn enables even more sophisticated attacks.

The mathematical foundation underlying these attacks treats the target system as an unknown function to be reverse-engineered through strategic sampling \cite{chen2017zoo,ilyas2018black}. Rather than computing exact internal gradients (which would require access to the model's architecture and parameters), attackers estimate gradient-like information by observing how scores change when they perturb inputs in different directions. By testing variations and using the score feedback to update their search strategy, attackers effectively learn the local geometry of the system's scoring function. This enables precise attack crafting even without any knowledge of how the system works internally—the numerical scores alone provide sufficient signal.

From an acquisition perspective, organizations must weigh operational and cost benefits against attack vulnerability when deploying score-based face recognition APIs. While many assume cloud-based commercial services provide security advantages, numerical score outputs that enable adaptive thresholds and quality assessment also enable efficient adversarial attacks. Organizations should conduct explicit threat modeling to determine whether score-based access is operationally necessary. For identity verification requiring only 1:1 matching, binary accept/reject decisions may suffice and should be preferred when possible.
When score-based APIs are necessary, mitigations include those from decision-only based systems. Additionally, quantized scores such as coarse-grained confidence categories (``low'', ``medium'', ``high'') rather than continuous scores, would make it much more difficult for an attacker to estimate the local gradient direction.

\subsection{Embedding: High-Dimensional Feature Vector Access}
In the Embedding access category, the adversary gains access to intermediate feature representations, typically vectors of 128 to 512 dimension, that encode facial identity information prior to final comparison or classification. This access level is increasingly prevalent in interoperable face recognition systems, software development kits (SDKs) that separate feature extraction from matching logic, and certain API architectures that expose embeddings to enable client-side comparison or downstream analytics. Embedding access represents a critical escalation in risk because these representations contain substantially more information than similarity scores, enabling both more effective evasion attacks and severe privacy violations through template inversion \cite{vakhshiteh2021adversarial}.

The threat expands along two dimensions: crafting adversarial examples and reconstructing face images from templates. For attacks, adversaries can optimize perturbations directly using the embedding vectors rather than working backward from scores. Given embeddings for both their own face (source) and a target victim's face, attackers formulate the problem as making the perturbed source embedding as close as possible to the target embedding while keeping the image changes imperceptible \cite{sharif2016accessorize,zhou2024rethinking,komkov2021advhat,wang2021similarity,zhong2020towards}. This formulation, minimizing feature distance between source and target while maintaining perceptual similarity, is the standard approach for targeted impersonation attacks when embedding access is available. This approach is more effective than score-based attacks because it operates in the representation space where the system actually compares identities, not just the final yes/no outcome.

Organizations acquiring face recognition systems should explicitly assess whether access to facial embeddings is functionally necessary, as this architectural decision directly determines the system's attack surface. If an attacker can access the embeddings, they can perform the same attacks they can perform with score-based APIs, and it likely reduces the number of queries required as they do not have to request matches between each photo (grows $N^2$), but rather retrieve an embedding for each photo (grows $N$).


\textbf{Related consideration: privacy implications of embedding access}
The privacy risks associated with access to embeddings are equally severe.
Model inversion attacks demonstrate that adversaries with access to face recognition models can reconstruct recognizable face images. Early work showed this is possible by exploiting confidence scores from classification models \cite{fredrikson2015model}, while more recent attacks target the embeddings themselves \cite{mai2018reconstruction}. Research on inverting face embeddings shows that reconstructed images can successfully impersonate enrolled identities, achieving high true accept rates on verification tasks and identification rates in gallery matching when tested against face recognition systems \cite{mai2018reconstruction}. This demonstrates that embeddings, despite being lower-dimensional representations, retain sufficient biometric information to enable practical identity recovery attacks.

For organizations storing face embeddings rather than raw images, a common practice motivated by privacy and storage efficiency, embedding access through data breach or insider threat still represents catastrophic identity compromise. Unlike passwords, biometric templates cannot be reset or reissued. Recent research demonstrates that attackers can train generative or diffusion models to invert embeddings: given an embedding vector, the generator produces a face image that, when passed back through the system, yields the same embedding \cite{mai2018reconstruction,shahreza2023comprehensive,shahreza2025face,wang2025diffusion,wang2025diffumi}. This effectively converts an embedding database breach into a complete biometric database breach.

\subsection{White-Box: Complete Model and Training Knowledge}
White-box access describes scenarios in which an adversary has full visibility into a face recognition system, including model architecture, trained parameters, preprocessing steps, and training procedures. This access level represents the most severe point in the taxonomy because it eliminates uncertainty about how the system behaves. Although white-box threats are often assumed to arise only from extreme failures or open-source deployments, multiple surveys emphasize that such access is increasingly realistic in practice due to public model repositories, outsourced development, insider access, edge deployment reverse engineering, and regulatory or audit-driven disclosure requirements \cite{biggio2018wild,chakraborty2018adversarial,vakhshiteh2021adversarial}.

White-box access fundamentally changes evasion from exploratory behavior to deterministic, gradient-driven optimization. When an adversary has access to model gradients, they can directly observe how infinitesimal changes to an input influence internal representations and final decisions. This gradient information allows attackers to iteratively refine inputs in the most effective direction for causing misrecognition, rather than relying on guesswork or repeated trial-and-error. Gradient availability is the defining advantage of white-box access, enabling attacks that are faster, more reliable, and require substantially smaller input modifications than black-box alternatives \cite{biggio2018wild,akhtar2018threat}.

From an acquisition perspective, white-box access is not a theoretical edge case but a plausible outcome of common design choices. Organizations that fine-tune publicly available face recognition models inherit white-box exposure by default. Contracted developers and system integrators necessarily operate with full model access, creating insider threat considerations. Edge deployments, such as mobile devices or embedded access-control systems, are often vulnerable to reverse engineering, effectively converting nominally black-box systems into white-box ones post-deployment \cite{biggio2018wild}.
The defensive literature is largely unified in its assessment that no defense can fully eliminate risk in a white-box setting. Instead, defenses aim to raise attack cost and reduce reliability. Researchers consistently identify adversarial training, randomized inference behavior, ensemble approaches, and adversarial detector techniques as the most effective available mitigations, while also noting their tradeoffs in accuracy, performance, and operational complexity \cite{cohen2019certified,madry2017towards,ren2020adversarial}. As a result, organizations should evaluate systems under white-box assumptions even when deployment plans assume restricted access, as model leakage frequently occurs after systems are operational.
While procuring face recognition systems, white-box resilience should be treated as a baseline evaluation criterion, not a worst-case afterthought. Systems that appear robust under limited-access testing may fail catastrophically once model details are exposed. Incorporating white-box evaluation into procurement requirements aligns system assessment with the realities documented across a decade of adversarial machine learning research.

\textbf{Related consideration: data poisoning}
White-box access also extends beyond test-time evasion to include training-time manipulation, such as data poisoning and backdoor insertion. When attackers have insight into, or influence over, the training process, they can implant failures that remain dormant under normal conditions and activate only in attacker-chosen scenarios \cite{papernot2016limitations,chakraborty2018adversarial}. These risks are especially relevant in collaborative training environments, reliance on externally sourced datasets, and procurement pipelines that involve multiple vendors or contractors.

\section{Risk Landscape Analysis and Procurement Decision Considerations}
\label{sec:decision}

The progression from None to White-Box access reveals a non-linear escalation in adversarial capabilities and attack efficiency. Understanding this landscape is essential for mapping procurement choices to threat models and making evidence-based security-utility tradeoffs.

\subsection{Query Efficiency Across Access Levels}
Attack query budgets vary dramatically across access categories. Attacks with None and Metadata access operate entirely offline, crafting adversarial examples without system interaction. Decision-Only attacks require extensive query budgets to probe decision boundaries. Score/Rank access enables gradient estimation with substantially reduced query requirements, often one to two orders of magnitude fewer than decision-based attacks. Embedding access provides direct optimization targets, while White-Box access eliminates queries entirely through offline gradient computation.
This progression reveals critical inflection points. The transition from None/Metadata to Decision-Only enables interactive attacks but imposes substantial query costs. The shift to Score/Rank dramatically reduces query budgets through gradient approximation. White-Box access eliminates system interaction requirements, enabling pre-computed attacks.

\subsection{Attack Success and Transferability}
Success rates escalate with access level, though not uniformly. Even None-access scenarios achieve non-trivial transfer rates against unseen systems, establishing a baseline risk independent of deployment architecture. Decision-Only attacks achieve high success with sufficient queries. Score/Rank access reaches near-perfect success with moderate query budgets. White-Box attacks approach certainty against the target system while maintaining substantial transfer rates to architecturally similar models.
The substantial jump in both success rates and query efficiency between Decision-Only and Score/Rank access underscores the security cost of continuous feedback. Organizations should recognize that transferability means successful attacks against one system often compromise similar deployments.

\subsection{Privacy Risk Escalation}
Privacy violations through model inversion and membership inference follow different escalation patterns. None, Metadata, and Decision-Only access pose limited privacy risk—attacks focus on evasion rather than reconstruction. Score/Rank access introduces moderate privacy risk through membership inference from confidence patterns. Embedding access creates severe privacy risk by enabling face reconstruction from stored templates. White-Box access represents critical privacy failure, allowing training data extraction and perfect inversion.
For organizations storing biometric templates, Embedding and White-Box categories could cause catastrophic privacy failures. Procurement decisions involving embedding-based architectures must account for data breach notification requirements, regulatory compliance costs, and reputational damage beyond direct evasion threats.

\subsection{Mapping Procurement Options to Access Categories}
Table \ref{tab:procurement-access-actors} illustrates how procurement decisions implicitly define the maximum plausible level of model access available to different classes of adversaries over a system’s lifetime. Rather than assuming a single attacker model, the table distinguishes between external attackers and insider vulnerabilities scenarios. While many deployments are designed to expose only limited signals, e.g., decision-only or score-based outputs, practical considerations such as insider access, outsourcing, employee turnover, edge deployment, and device compromise frequently elevate adversary access toward white-box conditions. The key insight for procurement is that access levels are not static: even systems initially deployed under restrictive assumptions may drift toward higher-access threat models over time. Consequently, procurement choices should be evaluated not only on their intended exposure, but on the worst-case access level they plausibly enable.

\begin{table}[!htbp]
\centering
\caption{Maximum Plausible Model Access by Procurement Option and Threat Actor}
\label{tab:procurement-access-actors}
\begin{tabularx}{\textwidth}{p{6cm}p{5cm}Y}
\toprule
\textbf{Procurement Option} &
\textbf{External Attacker} &
\textbf{Insider Threat} \\
\midrule
Open-source base models &
White-Box &
White-Box \\

Commercial cloud APIs &
Score/Rank &
Metadata to Embedding \\

Commercial on-premise systems &
Decision-Only to None &
Metadata to Embedding \\

Custom development (contracted) &
None to Decision-Only &
White-Box \\

In-house development &
None to Decision-Only &
White-Box \\

Edge deployment &
None to Decision-Only &
White-Box \\
\bottomrule
\end{tabularx}
\end{table}

Organizations should approach AI system procurement using a structured framework. Such a framework should incorporate the five following steps. First, define relevant adversary classes (opportunistic, targeted, insider), estimate their resources and time horizon, assess attack value, and identify applicable privacy and regulatory constraints. Second, map realistic access levels for each procurement option across external attackers, insiders, and post-compromise scenarios, using Table \ref{tab:procurement-access-actors} as a baseline adjusted for organizational controls. Third, evaluate attack feasibility by considering adversary capability, privacy exposure, regulatory impact, operational requirements, and total cost of ownership, including monitoring, incident response, compliance, and breach remediation. Fourth, align defenses to access level: None/Metadata require liveness checks and rate limiting; Decision-Only adds query logging and lockouts; Score/Rank adds output control and model-stealing detection; Embedding requires strong template protection and breach response; White-Box demands adversarial training, architectural diversity, and rigorous robustness testing. Fifth, codify these requirements in procurement specifications, including adversarial evaluation criteria, monitoring and notification obligations, access controls, and model update policies.

\section*{Declarations}

\subsection*{Competing Interests}
On behalf of all authors, the corresponding author states that there is no conflict of interest.

\subsection*{Funding Information}
This research was funded under Interagency Agreement 70RSAT23KPM000049 by the U.S. Department of Homeland Security (DHS) Science and Technology (S\&T) Directorate.

\subsection*{Author contribution}
Maria Mahbub, Steven Young, and Amir Sadovnik conceived the idea for the survey. Maria Mahbub conducted the literature review, analyzed the existing studies, developed the taxonomy, and drafted the manuscript. All authors critically revised the manuscript, provided intellectual input, reviewed, and approved the final manuscript.

\subsection*{Data Availability Statement}
Not Applicable

\subsection*{Research Involving Human and /or Animals}
Not Applicable

\subsection*{Informed Consent}
Not Applicable

\bibliographystyle{splncs04}
\bibliography{ref} 

\end{document}